# Unraveling the spin reorientation process in rare earth perovskite $PrFe_{0.1}Cr_{0.9}O_3$

Jiyu Shen[1], Jiajun Mo[1], Zeyi Lu[1], Chenying Gong[1], Zongjin Wu[1], Kaiyang Gao[1], Min Liu*,[1], Yanfang Xia*,[1]

[1]College of Nuclear Science and Technology, University of South China, Hengyang 421200, Hunan, P.R China.

*Email: liuhart@126.com, xiayfusc@126.com

**Abstract**

Ultrafast spin control plays a pivotal role in condensed matter physics. In this study, we analyzed the macroscopic magnetization of the $PrFe_{0.1}Cr_{0.9}O_3$ system by molecular field model fitting. And the whole process of system spin reorientation is accurately calculated in the fitting process. It is found that, unlike the rare-earth perovskites we have previously studied, $PrFe_{0.1}Cr_{0.9}O_3$ exhibits spin-reversion properties during the reorientation process. This research will lay a theoretical foundation for precise spin control in the future.

**Introduction**

Praseodymium-based rare earth perovskites are widely used in many fields. Tang et al. [1] studied the ability of lanthanum-praseodymium-based double perovskites to catalyze the degradation of gaseous toluene; Liu et al. [2] studied the electrochemical performance of praseodymium-neodymium-based perovskites; Chen et al. [3] investigated the performance changes of photoluminescence induced by $Pr^{3+}$ doping.

Not only that, but there is a very attractive physical phenomenon in praseodymium-based rare earth perovskites, namely spin reorientation. Nowadays, spin switch technology has appeared in many high-end technology products. But in order to precisely control the spin properties of ions, enormous efforts are often required. Currently known and widely used control methods are nothing more than femtosecond laser control [4-6] and thermal control [7-10]. Generally speaking, the high consumption of femtosecond laser control and the low sensitivity of thermal control greatly limit the development of spin control technology. How to reduce the consumption of laser control or how to improve the accuracy of thermal control has become an important work in condensed matter physics.

In this study, we will conduct a detailed study of the magnetic characteristics of $PrFe_{0.1}Cr_{0.9}O_3$. We will determine the macroscopic magnetic characteristics at fixed temperature by means of Mössbauer spectra and magnetization curves. The thermomagnetic curves were fitted by the molecular field model, and the specific exchange constants were obtained. Most importantly, we will precisely reveal the complete process of $PrFe_{0.1}Cr_{0.9}O_3$ spin reorientation during the fitting process, and make corresponding conjectures. This work will provide theoretical guidance for spin control of praseodymium-based rare earth perovskites.

**Experimental details**

Sample Preparation

PrFe$_{0.1}$Cr$_{0.9}$O$_3$ was prepared by a simple sol-gel combustion method. The required precursors are all from McLean, namely praseodymium nitrate hexahydrate (Pr(NO$_3$)$_3$ 6H$_2$O), iron nitrate nonahydrate (Fe(NO$_3$)$_3$ 9H$_2$O), chromium nitrate nonahydrate (Cr(NO$_3$)$_3$ 9H$_2$O), ethylene glycol (C$_2$H$_6$O$_2$), citric acid (C$_6$H$_8$O$_7$). First, all nitric acid compounds were mixed and dissolved in deionized water according to a certain ion ratio (Pr$^{3+}$: Fe$^{3+}$: Cr$^{3+}$=1: 0.1: 0.9). Then, the excess citric acid and a certain amount of ethylene glycol are mixed in a molar ratio of 1:1. Stir the solution uniformly on a magnetic stirrer at 80 °C until a gel forms. Then, heat the gel to 120°C. The resulting powder was pre-calcined at 600°C for 12 hours, then calcined at 1200°C for 24 hours, and cooled to obtain the final nanopowder sample.

X-ray diffraction measurements (XRD)

X-ray diffraction (XRD) experiments were carried out in Siemens D500 Cu Kα (λ = 1.5418 Å) diffractometer with the range of 20° to 80° (rate of 0.02°/s). The obtained data were processed with Fullprof software.

Mössbauer spectrum Test

The transmission $^{57}$Fe Mössbauer spectra of PrFe$_{0.1}$Cr$_{0.9}$O$_3$ were collected at RT on SEE Co W304 Mössbauer spectrometer with a $^{57}$Co/Rh source in transmission geometry equipped in a cryostat (Advanced Research Systems, Inc.,4 K). The data results were fitted with MössWinn 4.0 software.

Magnetic Test

The thermomagnetic curves (M-T) of the samples $PrFe_{0.1}Cr_{0.9}O_3$ were obtained under an external magnetic field of 100 Oe. Field-cooled (FC) and zero-field-cooled (ZFC) measurements were done at a magnetic field strength of 100 Oe and temperatures ranging from 5 to 400 K. The magnetization curves (M-H) are obtained at temperatures of 300 K and 5 K, respectively.

**Results and discussions**

XRD

analysis

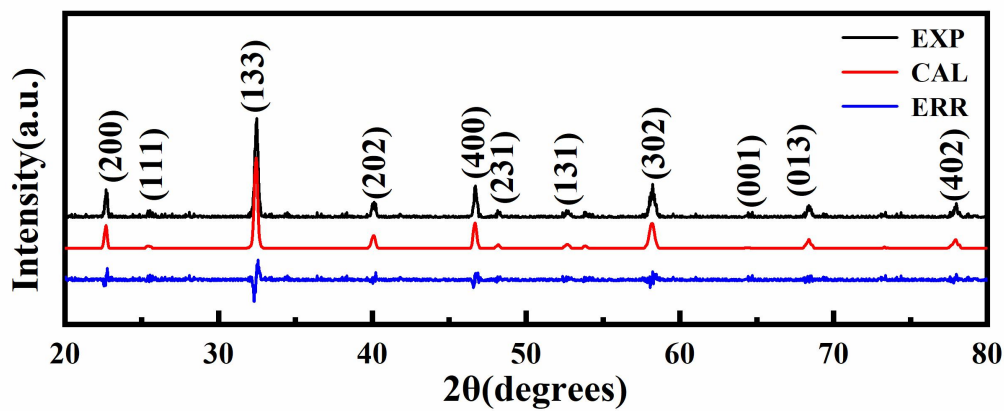

**Fig. 1.** X-ray diffraction pattern of $PrFe_{0.1}Cr_{0.9}O_3$, (the main diffraction Bragg peaks are marked in the figure).

The X-ray diffraction spectrum of the $PrFe_{0.1}Cr_{0.9}O_3$ perovskite sample at room temperature was obtained and refined with the Fullprof software. The results are shown in Figure 1. By comparing the angular positions of the diffraction peaks, we can confirm that the sample belongs to the orthorhombic single-phase structure of *Pbnm*, and no impurity phase is observed. This is the general structure of most praseodymium-based perovskites [11-13].

The lattice parameters etc. obtained by the refinement of Fullprof software are shown in Table 1. Compared with previous studies [11,12], we found that after doping with 10% Fe, there is a small increase in lattice parameters, which is very common and is related to the radii of $Fe^{3+}$ and $Cr^{3+}$. During the refinement process, we estimated the antisite defect factor ASD of this sample to be about 0%, which means that in this system, $Fe^{3+}$ ions and $Cr^{3+}$ ions are completely ordered. Combined with our previous theory, we put this The value of ASD is converted to x = 5.99999, y = 4.80001, these two parameters will play a very important role in the subsequent process of fitting the magnetism [14].

Table. 1 Lattice parameters, unit cell volume and other parameters of $PrFe_{0.1}Cr_{0.9}O_3$.

| Sample | a (Å) | b (Å) | c (Å) | Cell volume (Å³) | Average grain size (Å) | Bulk density (g/cm3) |
|---|---|---|---|---|---|---|
| $PrFe_{0.1}Cr_{0.9}O_3$ | 5.514 | 5.487 | 7.752 | 58.86 | 376 | 6.7953 |

Magnetization curve (M-H) and Mössbauer analysis

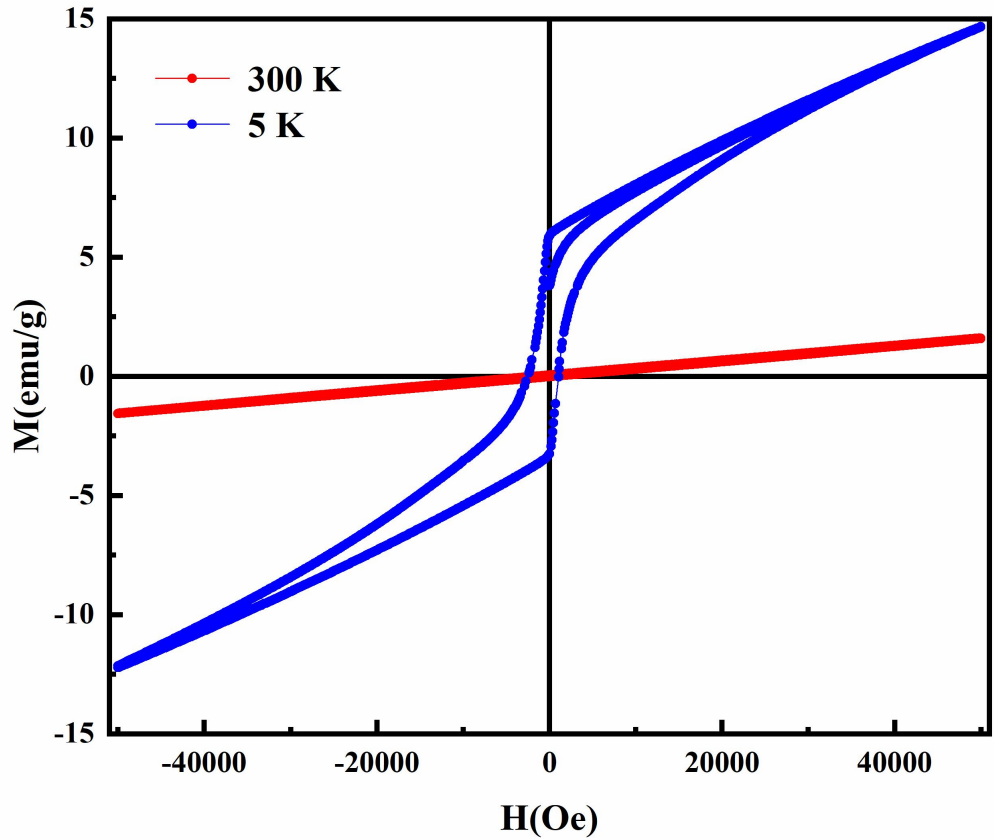

**Fig. 2.** Magnetization curves of PrFe$_{0.1}$Cr$_{0.9}$O$_3$ at 300 K and 5 K, respectively.

In order to obtain the macroscopic magnetism at a fixed temperature, we tested its magnetization curves at 300 K and 5 K, respectively, and the results are shown in Figure. 2. Undoubtedly, at 5 K, the sample exhibits a very pronounced ferromagnetic coupling phenomenon. This can be understood due to the superexchange interaction (Fe$^{3+}$/Cr$^{3+}$-O-Fe$^{3+}$/Cr$^{3+}$) that arises by the introduced Fe$^{3+}$ ions, i.e., there are four groups of antiparallel spin direction in the Fe/CrO$_6$ octahedral structure containing Fe$^{3+}$/Cr$^{3+}$-O$^{2-}$. Consequently, the remaining electron and spin-orbit coupling induces superexchange interaction facilitating the spin angle of the Fe$^{3+}$/Cr$^{3+}$ ion reduction to slightly less than 180°. At this time, the net magnetic moment at the B site is not zero and exhibits ferromagnetic behavior under the action of an external field [15].

The macroscopic magnetism at room temperature may exhibit strong paramagnetism or strong antiferromagnetism. In order to determine its true macroscopic magnetic characteristics, we tested its Mössbauer spectrum at 300 K, and the results are shown in Figure. 3.

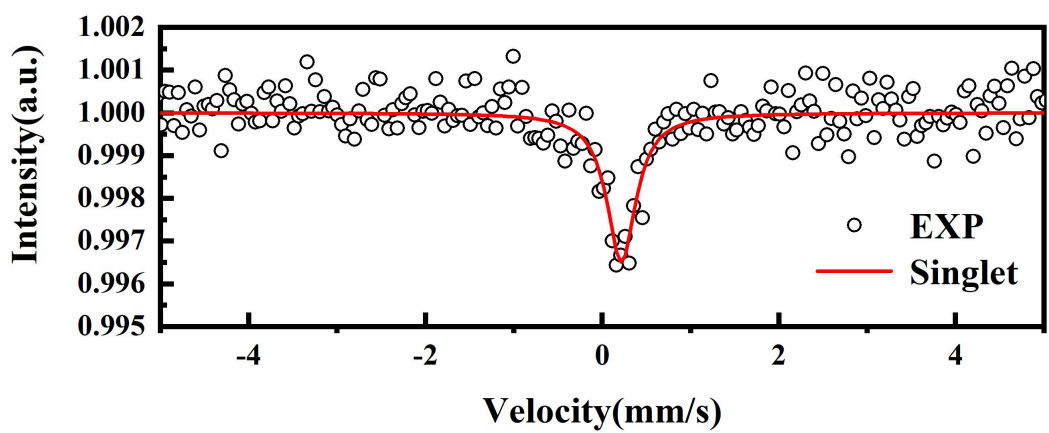

**Fig. 3.** Mössbauer spectrum of $PrFe_{0.1}Cr_{0.9}O_3$ at 300 K

The fitted spectrum appears as a single singlet feature with no other components, especially sextets. Therefore, we can conclude that the sample has undergone the phase transition at the Neel temperature at 300 K, which belongs to the paramagnetic property. This also shows that 10% Fe doping does not make the Neel temperature point of the sample exceed 300 K, which is very close to the previous study [16].

In addition to this, we found a very interesting phenomenon in the M-H curve: the wasp waist phenomenon. In a general sense, the formation of the wasp waist hysteresis loop is mainly due to the joint action of two or more phases of magnetic phases with different coercive forces [17-19]. This has also been mentioned in previous studies. In particular, Jiashu Zhang et al.'s [20] report on the discovery of wasp waist hysteresis loops in rare-earth perovskites for the first time gives us reason to believe that the sample is ultimately caused by the combined action of antiferromagnetic and ferromagnetic components. The antiferromagnetic component at 5 K may be due to the electric field gradient distortion caused by the large QS, which makes it difficult for $Fe^{3+}/Cr^{3+}$ ions to form a small angle and keep the antiferromagnetic coupling all the time.

**Thermomagnetic Curve Analysis and Simulation**

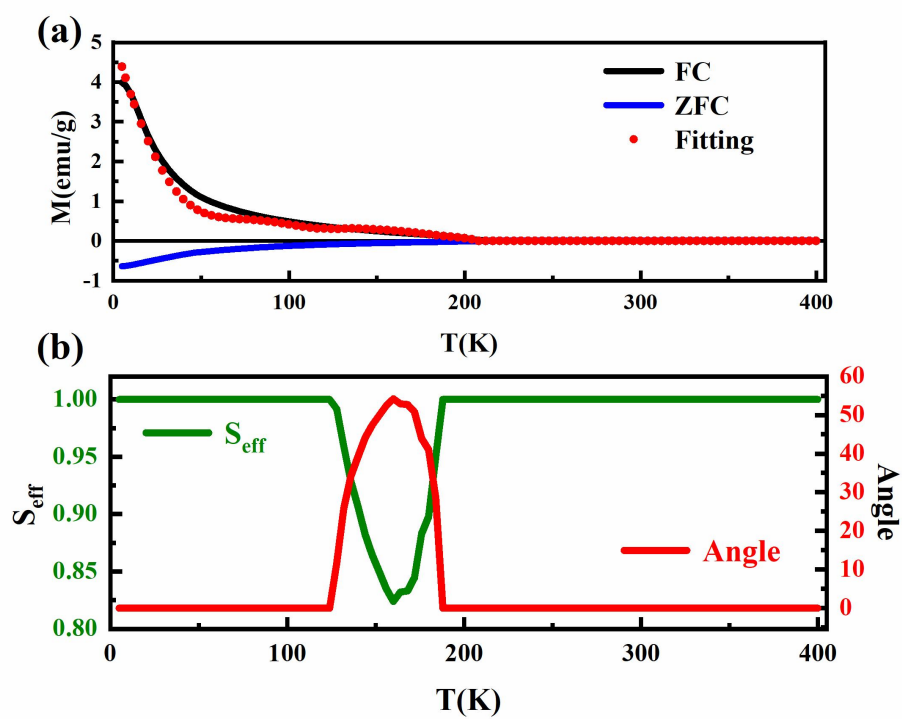

**Fig. 4.** Thermomagnetic curve data and fitting results of $PrFe_{0.1}Cr_{0.9}O_3$. (a) The M-T experimental data of $PrFe_{0.1}Cr_{0.9}O_3$ and the fitting result of the FC curve, (b) the fitting result of the spin reorientation process of $PrFe_{0.1}Cr_{0.9}O_3$, the left side shows the change of the effective magnetic moment, and the right side shows the change of the included angle.

The experimental data of the thermomagnetic curves (M-T) of $PrFe_{0.1}Cr_{0.9}O_3$ are presented in Figure. 4(a). We can find that after about 210 K, the field cooling curve tends to the abscissa axis, indicating that the sample is undergoing a phase transition at 210 K. Compared with our previously studied $NdFe_{1-x}Cr_xO_3$ system [21], the spin reorientation temperature of $PrFe_{0.1}Cr_{0.9}O_3$ is difficult to see from the M-T curve of macroscopic magnetism. Therefore, for such materials, it is very difficult to explore the spin reorientation process, especially the precise positioning.

We consider a relatively straightforward method to reveal the magnetic properties of a system in order to better understand the magnetic mechanism, namely the molecular field theory. $PrFe_{0.1}Cr_{0.9}O_3$ is characterized by four sublattices, denoted by the letters $L_{Fe}$, $L_{Cr}^a$, $L_{Cr}^b$, and $L_{Pr}$, which correspond to the sublattices decorated by the $Cr^{3+}$ ions in the a- and b-sites, respectively. In this work, we do not consider the Pr-Pr superexchange because it has a very low phase transition temperature. So it is too insignificant to affect the system. The mean-field of each sublattice can be expressed as follows:

$$H_{Fe} = \lambda_{FeCr^a} M_{Cr^a} + \lambda_{FeNd} M_{Nd} + h$$
$$H_{Cr^b} = \lambda_{Cr^bCr^a} M_{Cr^a} + \lambda_{Cr^bNd} M_{Nd} + h$$
$$H_{Cr^a} = \lambda_{Cr^aCr^b} M_{Cr^b} + \lambda_{Cr^aFe} M_{Fe} + \lambda_{Cr^aNd} M_{Nd} + h \quad (1)$$
$$H_{Pr} = \lambda_{PrFe} M_{Fe} + \lambda_{PrCr^b} M_{Cr^b} + \lambda_{PrCr^a} M_{Cr^a} + h$$

Where $\lambda_{ij}$ represents the molecular field constant between $i$ and $j$ sublattices, and it is proportional to the exchange constant $J_{ij}$, $h$ is the external field; $M_i$ is the magnetization of $i$ sublattice. The magnetization of i sublattice is:

$$M_i = x_i N_A g \mu_B S_i B_{Si}\left(\frac{g\mu_B S_i H_i}{k_B T}\right) \quad (2)$$

Where $x_i$ is the molar quantity of $i$ ions, $g$ is the lande factor, and $\mu_B$ represents the Bohr magneton. $N_A$ is the Avogadro constant. $S_i$ is the spin quantum number of $i$ ions ($S_{Fe}$ = 5/2, $S_{Cr}$ = 3/2, $S_{Pr}$ = 1). The exchange constant between two $M$ ions may be determined by the following formula:

$$|J_{MM}| = \frac{2Z_{MM} S_M (S_M + 1)}{3 k_B T_N^M} \quad (3)$$

Herein $Z_{MM}$ is the number of $M$ ions required to be $M$ ions nearest neighbours. $k_B$ is the Boltzmann constant, and $T_N^M$ is the phase transition temperature of PrMO$_3$. The exchange constants between Fe and Fe ($J_{FeFe}$) and Cr and Cr ($J_{CrCr}$) have been calculated to be 14.67 K and 20.0 K ($T_N$ of PrCrO$_3$~220 K, $T_N$ of PrFeO$_3$ ~ 702 K). Using Eqs. (1) and (2) concurrently, the magnetization at each temperature can be calculated (3). To fit the experimental data using molecular field theory, we employ the most well-known heuristic algorithm—the Marine Predator Algorithm (MPA) [22].

Due to the fact that the angle between ions varies with the ratio of $Cr^{3+}/Fe^{3+}$, it is reasonable to anticipate that the exchange coupling constant will also vary. Using the assumption that the exchange rate has changed by no more than 10%, an exact exchange rate between $0.9J_{ij}$ and $1.1J_{ij}$ can be determined. Thus, the parameters $J_{PrFe}$, $J_{PrCr}$, and $J_{FeCr}$ are determined using the best fit of the experimental curve. Although we have performed numerous calculations to find appropriate parameters, the fitting result does not capture the entire picture. Indeed, molecular field theory is an oversimplified model that ignores Dzyaloshinskii-Moriya (DM) interactions, spin redirection, or anisotropy. We consider the magnetism of $PrFe_{0.1}Cr_{0.9}O_3$ as a vector superposition of $Fe^{3+}/Cr^{3+}$ and $Pr^{3+}$ ions to demonstrate the effect of spin reorientation on magnetism. As A and B sites have different orientations of their easy axes, their interaction can be determined by their spin projection on a particular plane.

The final fitting result is shown in Figure. 4(a). Comparing the experimental data, we see that the accuracy of the fitting is good. The corresponding exchange constants obtained can be found in Table 2. In the fitting process, considering the effect of spin reorientation, we define the projection of the magnetic moment of the $Pr^{3+}$ ion on the $Fe^{3+}/Cr^{3+}$ magnetic moment plane as the effective magnetic moment. The fitting result is shown in Figure. 4(b). It is found that there is a symmetrical peak in the variation of the effective magnetic moment. This is also quite different from our previous study of $NdFe_{1-x}Cr_xO_3$ [21]. It shows that in the process of this redirection, the change of moment satisfies a back-and-forth oscillation, and the corresponding angle change also returns after reaching 55°. For a simple process model, please refer to Figure 5.

**Table. 2.** Exchange constants for $PrFe_{0.1}Cr_{0.9}O_3$ systems.

| $J_{Pr-Fe}$ (K) | $J_{Pr-Cr}$ (K) | $J_{Fe-Fe}$ (K) | $J_{Cr-Cr}$ (K) | $J_{Fe-Cr}$ (K) |
|---|---|---|---|---|
| -2.000 | -3.112 | -28 | 14.378 | -4.922 |

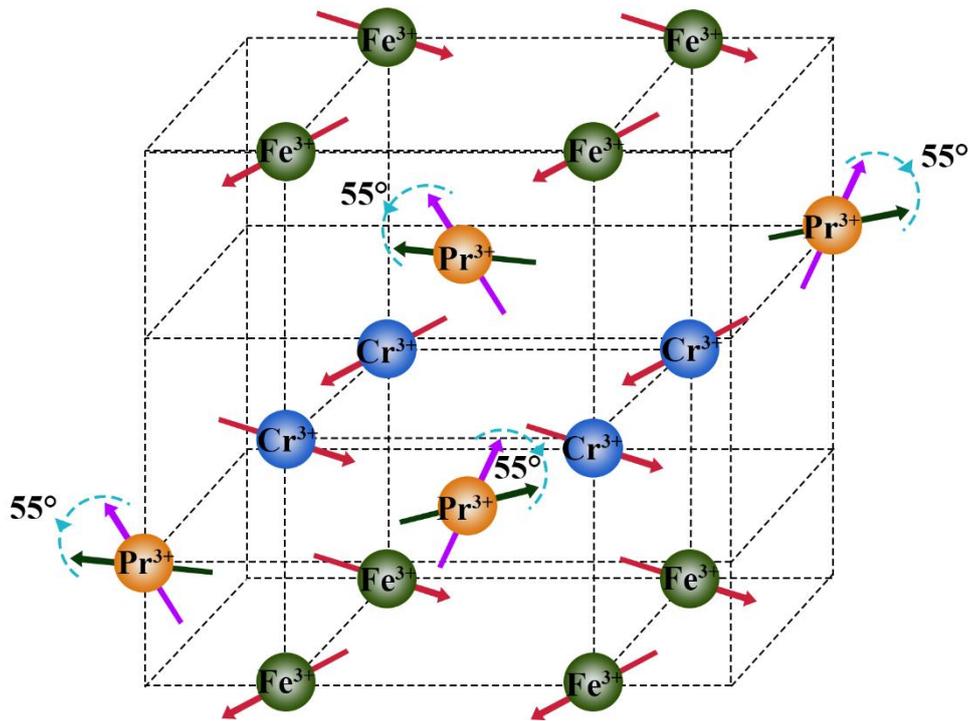

**Fig. 5.** Model of the spin reorientation process for PrFe$_{0.1}$Cr$_{0.9}$O$_3$.

**Conclusions**

In summary, we synthesized a single-phase orthogonal PrFe$_{0.1}$Cr$_{0.9}$O$_3$ system by a simple sol-gel method. The macroscopic magnetic characteristics at 300 K were determined by magnetization curves and Mössbauer spectra. The thermomagnetic curves of the samples were fitted by the molecular field model, and the results were satisfactory. And in the fitting process, the effective magnetic moment is defined as the projection of the Pr$^{3+}$ magnetic moment on the Fe$^{3+}$/Cr$^{3+}$ magnetic moment plane, thus the complete process of its spin reorientation is obtained. This study provides theoretical guidance for spin control of praseodymium-based rare earth perovskites.

**Conflicts of interest**

There are no conflicts to declare.

**Acknowledgements**

This work was supported by National Natural Science Foundation of China (grant number 12105137, 62004143), the Central Government Guided Local Science and Technology Development Special Fund Project (2020ZYYD033), the National Undergraduate Innovation and Entrepreneurship Training Program Support Projects of China, the Natural Science Foundation of Hunan Province, China (grant number S202110555177), the Natural Science Foundation of Hunan Province, China (grant number 2020JJ4517), Research Foundation of Education Bureau of Hunan Province, China (grant number 19A433, 19C1621).